\documentclass[12pt]{iopart}
\usepackage{amssymb}
\usepackage{setstack}
\usepackage{verbatim}
\usepackage{bm}
\usepackage{graphicx}

\newcommand{\bq}{\begin{equation}}
\newcommand{\eq}{\end{equation}}
\newcommand{\bn}{\begin{eqnarray}}
\newcommand{\en}{\end{eqnarray}}

\begin{document}

\newcommand{\cc}{{\bf\Large C }}
\newcommand{\hide}[1]{}
\newcommand{\tbox}[1]{\mbox{\tiny #1}}
\newcommand{\half}{\mbox{\small $\frac{1}{2}$}}
\newcommand{\sinc}{\mbox{sinc}}
\newcommand{\const}{\mbox{const}}
\newcommand{\trc}{\mbox{trace}}
\newcommand{\intt}{\int\!\!\!\!\int }
\newcommand{\ointt}{\int\!\!\!\!\int\!\!\!\!\!\circ\ }
\newcommand{\eexp}{\mbox{e}^}
\newcommand{\EPS} {\mbox{\LARGE $\epsilon$}}
\newcommand{\ar}{\mathsf r}
\newcommand{\re}{\mbox{Re}}
\newcommand{\bmsf}[1]{\bm{\mathsf{#1}}}
\newcommand{\dd}[1]{\:\mbox{d}#1}
\newcommand{\abs}[1]{\left|#1\right|}
\newcommand{\ket}[1]{| #1 \rangle}
\newcommand{\bra}[1]{\langle #1 |}
\newcommand{\mbf}[1]{{\mathbf #1}}

\title{Cavity-induced giant Kerr nonlinearities in a driven $V$-type atom}
\author{Rong Tan$^{1,2}$, Gao-xiang Li$^{1}$ and Zbigniew Ficek$^{3,4}$}
\eads{\mailto{gaox@phy.ccnu.edu.cn}}
\address{$^{1}$Department of Physics, Huazhong Normal University, Wuhan 430079, China}
\address{$^{2}$College of Science, Wuhan Institute of Technology, Wuhan 430073, China}
\address{$^{3}$Department of Physics, School of Physical Sciences, The University of Queensland,
Brisbane, Australia 4072}
\address{$^{4}$The National Centre for Mathematics and Physics, KACST, P.O. Box 6086, Riyadh 11442, Saudi Arabia}

\date{today}

\begin{abstract}
We discuss a simple and experimentally realizable model for creation of enhanced Kerr nonlinearities accompanied by vanishing absorption. The model involves a $V$-type atom subjected to a strong drive laser, a weak probe laser and coupled to a single-mode cavity field. Working in the bad-cavity limit, we find that the simultaneous coupling of the cavity field to both atomic transitions creates a coherence between the transitions and thus can lead to quantum interference effects.
We investigate the influences of the cavity field frequency, the cavity field-atom coupling constants and the atomic decay constants on the linear and the third-order (Kerr) nonlinear susceptibilities. We predict giant Kerr nonlinearities with vanishing absorption and attribute this effect to the combination of the Purcell effect and the cavity-induced quantum interference.
\end{abstract}

\maketitle

\section{Introduction}

A great deal of attention has been focused recently on the creation of strong nonlinear effects in single coherently prepared multi-level atoms~\cite{hh,fs02}. The motivation for this interest is to fabricate an atomic medium with giant nonlinear properties produced with relatively low light powers. A particular attention has been paid to the third-order Kerr-type nonlinearities which play an important role in nonlinear optics and have many fascinating  applications in different areas of physics ranging from phase modulation~\cite{Schmidt}, generation of optical solitons~\cite{V}, optical switching~\cite{os} to optical communication and computing~\cite{co}. Important for practical applications is to achieve enhanced or giant Kerr nonlinearities in an atomic medium with significantly reduced or even completely cancelled absorption rate for the propagating light beam. Imamo\v{g}lu {\it et al.}~\cite{Werner} have proposed a scheme to produce giant Kerr nonlinearities together with reduced absorption, by using
quantum interference effects related to electromagnetically induced transparency. In a four-level double-dark resonance system, Kerr nonlinearity can be enhanced several orders of magnitude accompanied
by vanishing linear absorption under the condition of the effective interaction of double dark
resonances~\cite{Niu1}. A number of different atomic schemes have been suggested to achieve a large nonlinearity with vanishing absorption~\cite{Matsko,Naka}. More recently, Niu and Gong~\cite{Niu} and Yan {\it et al.}~\cite{Yan} have shown that the Kerr nonlinearity can be enhanced with vanishing linear and nonlinear absorptions due to the spontaneously generated coherence~\cite{Zhu}.

The major obstacle in experimental investigations of the nonlinear
properties of multi-level atoms is the difficulty to find suitable
systems to create quantum interference effects between atomic
transitions responsible for the cancelation of the absorption of a
propagating field. Most of the schemes proposed have assumed that
the quantum interference occurs between two transitions with
parallel or anti-parallel dipole moments. In atoms with quantum
states close in energy the dipole moments are usually perpendicular.
Therefore, several schemes have been suggested to engineer quantum
interference effects in atoms with perpendicular dipole moments.
Most of the schemes suggests to use single-mode optical cavities
with preselected polarization in bad cavity limit~
\cite{L,gaox,Zhou}. Bermel {\it et al.}~\cite{Bermel} have found
that the Purcell effect \cite{p} can substantially influence the
Kerr nonlinearity. Brand\~{a}o {\it et al.}~\cite{Brand} proposed a
method to produce self- and cross-Kerr photonic nonlinearities using
light induced Stark shifts arising from the interaction of a cavity
mode with atoms. In addition to the Purcell effect which is
substantial in optical cavities, where spontaneous transitions occur
only at selected frequencies, a cavity-induced quantum interference
is expected to arise which is analog of the spontaneously generated
interference~\cite{fs04}. Thus, a question arises, to what extent a
combination of the Purcell effect and the cavity-induced
interference effects will affect the susceptibility of driven
$V$-type three-level atom. The purpose of this paper is to address
this question and discuss in detail the possibility of obtaining
giant Kerr nonlinearities.

We consider a three-level atom in the $V$ configuration in which one of the two dipole allowed transitions is driven by a strong laser field while the other is probed by a weak beam. The atomic transitions are simultaneously coupled to a tunable single-mode cavity. Our interest will be centered principally on the effect of the cavity on the third-order susceptibility and determine if the driven system possesses enhanced or giant nonlinearities accompanied by vanishing absorption. The paper is organized as follows. In Section~\ref{ch2}, we introduce our model and outline the major steps in the derivation of the equations of motion for the density matrix elements. The iterative analytical solution for the coherences determining the susceptibility is presented in Section~\ref{ch3}. The results are presented graphically and discussed in Section~\ref{ch4}. We show the influences of the cavity field frequency, the cavity field-atom coupling constants and the atomic decay constants on the real and imaginary parts of the linear and nonlinear susceptibilities. We summarize our results in Section~\ref{ch5}.

\section{Theoretical Model}\label{ch2}

\begin{figure}[h]
\begin{center}
\includegraphics[width=8cm]{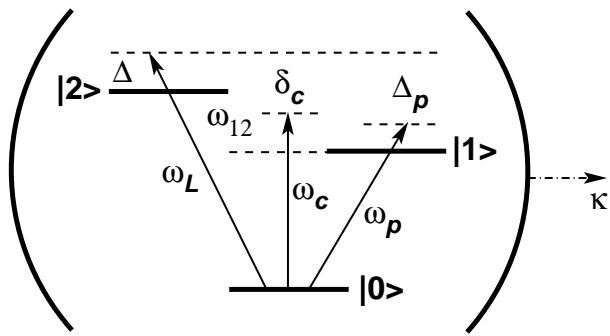}
\end{center}
\caption{Schematic diagram of the system. A three-level atom is located inside a single mode cavity strongly damped with a rate $\kappa$. The atomic transition $\ket 0 \leftrightarrow \ket 2$ is driven by a strong laser field of frequency $\omega_{L}$ and is probed by a weak laser field of a tunable frequency $\omega_{p}$ coupled to the $\ket 0\leftrightarrow \ket 1$ transition. Each of the two laser fields couples only to one of the atomic dipole transitions, while the cavity field couples to both transitions.}
\label{fig1}
\end{figure}

We consider a $V$-type three-level atom composed of two excited states $|1\rangle$
and $|2\rangle$ coupled to a common ground state $|0\rangle$ by transition dipole moments
$\vec{\mu}_{10}$ and $\vec{\mu}_{20}$, respectively. The atom is located inside a single-mode
cavity field of frequency $\omega_{c}$ and polarization~$\vec{e}_{c}$, as shown in Fig.~\ref{fig1}.
The polarization $\vec{e}_{c}$ is chosen such that the cavity field
is simultaneously coupled to both atomic transitions with the
coupling strengths $g_{1}$ and $g_{2}$, respectively. The atomic
transition $|2\rangle \rightarrow |0\rangle$ is driven by a strong
laser field of frequency $\omega_{L}$, whereas the $|1\rangle
\rightarrow |0\rangle$ transition is probed by a weak tunable laser
beam of frequency $\omega_{p}$. The cavity mode is damped at the
rate $\kappa$, whereas the atomic transitions are damped by
spontaneous emission to the modes other than the cavity mode at the
rates $\gamma_{1}$ and $\gamma_{2}$, respectively. In a frame
rotating at the frequency $\omega_{L}$, the master equation of the
density operator $\rho_{T}$ of the overall system (the atom and the
cavity field) is of the form
\begin{eqnarray}
\dot{\rho_{T}} = -i[H_{a}+H_{c}+H_{I},\rho_{T}]+{\cal L}_{a}{\rho_{T}}+{\cal L}_{c}{\rho_{T}} ,\label{e1}
\end{eqnarray}
where
\begin{eqnarray}
H_{a} &=& \Delta A_{22} -(\omega_{21}-\Delta)A_{11}+\Omega_{L}(A_{02}+A_{20})\nonumber \\
&& +\Omega_{p}e^{i\Delta_{p}t}A_{01}+\Omega_{p}e^{-i\Delta_{p}t}A_{10}
\end{eqnarray}
is the  unperturbed Hamiltonian of the coherently driven and weakly probed atom,
\begin{eqnarray}
H_{c} = \delta_{c} a^{\dag}a
\end{eqnarray}
is the Hamiltonian of the cavity field,
\begin{eqnarray}
H_{I} = g_{1}(a^{\dag}A_{01}+A_{10}{a}) + g_{2}(a^{\dag}A_{02}+A_{20}a)
\end{eqnarray}
is the interaction Hamiltonian of the cavity field with the atomic transitions, and
\begin{eqnarray}
{\cal L}_{a}\rho_{T} &=& \gamma_{1}(2A_{01}\rho_{T}A_{10}-A_{11}\rho_{T}-\rho_{T}A_{11})
\nonumber \\
&& +\gamma_{2}(2A_{02}\rho_{T}A_{20}-A_{22}\rho_{T}-\rho_{T}A_{22})\nonumber\\
&&+\gamma_{12}(2A_{01}\rho_{T}A_{20}-A_{21}\rho_{T}-\rho_{T}A_{21})
\nonumber \\
&& +\gamma_{12}(2A_{02}\rho_{T}A_{10}-A_{12}\rho_{T}-\rho_{T}A_{12}),\nonumber\\
{\cal L}_{c}\rho_{T} &=& \kappa(2a\rho_{T}a^{\dag}-a^{\dag}{a}\rho_{T}-\rho_{T}a^{\dag}{a})
\end{eqnarray}
are dissipative terms describing the damping of the atomic
transitions by spontaneous emission and of the field by cavity
decay.

Here, $a$ and $a^{\dag}$ are the annihilation and creation operators for the cavity field,
$A_{lk}=|l\rangle\langle{k}|\ (l,k=0,1,2)$ are the atomic operators,
$\omega_{21}=\omega_{2}-\omega_{1}$ is the frequency difference between the atomic transitions,
$\Delta=\omega_{2}-\omega_{L}$, $\Delta_{p}=\omega_{p}-\omega_{L}$ and
$\delta_{c} =\omega_{c}-\omega_{L}$ are the detunings of the atomic frequency $\omega_{2}$, the probe beam frequency~$\omega_{p}$ and the cavity frequency $\omega_{c}$ from the driving laser frequency $\omega_{L}$. The parameters, $\Omega_{L}=\vec{\mu}_{20}\cdot \vec{E}_{L}/\hbar$ and
$\Omega_{p}=\vec{\mu}_{10}\cdot\vec{E}_{p}/\hbar$ are the (real) Rabi frequencies of the driving laser field of amplitude $E_{L}$  and of the probe beam of amplitude $E_{p}$.

In writing the master equation (\ref{e1}), we have assumed that the
atomic dipole moments are not orthogonal to each other, which
results in the cross damping terms between the atomic transitions.
These terms lead to quantum interference between the two transitions
and are determined by the so-called cross damping parameter
$\gamma_{12}=\sqrt{\gamma_{1}\gamma_{2}}\cos\theta$, where $\theta$
is the angle between $\vec{\mu}_{10}$ and $\vec{\mu}_{20}$. When the
dipole moments are parallel ($\theta$ =0), the cross damping
parameter is maximal with $\gamma_{12}=\sqrt{\gamma_{1}\gamma_{2}}$,
whilst~$\gamma_{12}=0$ when the dipole moments are
perpendicular~\cite{fs04}. Quantum interference has been studied
intensively over years and has revealed new phenomena of both
conceptual and practical importance. It has been shown that
interference between atomic transitions induced by external fields
or by spontaneously created atomic coherence can lead to novel
phenomena such as electromagnetically induced transparency, lasing
without inversion, enhanced index of refraction and also nonlinear
processes such as enhanced the Kerr nonlinearities. However, most of
the predicted quantum interference effects have so far eluded
observation, as it is very unlikely to find isolated atoms with two
non-orthogonal dipole moments and states close in energy. Therefore,
we propose an alternative scheme where one can engineer coherence
between atomic transitions with perpendicular dipole moments by
coupling the transitions to a single-mode cavity field. As we shall
see, crucial for the creation of the coherence is to couple the
cavity mode simultaneously to both of the atomic transitions. In
practice, it can be easily achieved by setting a cavity-field
polarization making, for example, an angle $\alpha$ with the
direction of the atomic dipole moment $\vec{\mu}_{10}$ and that
simultaneously forms the angle $90^{\circ}-\alpha$ with the dipole
moment $\vec{\mu}_{20}$.

The master equation (\ref{e1}) we have started with is written in
the basis of the atomic states. Since the atomic transition
$|2\rangle\rightarrow |0\rangle$ is driven by a strong laser field,
it prompts us to introduce dressed states which provide a good
approach for studying the problem. The dressed states are
eigenstates of the Hamiltonian $H_{a}$ and are defined by the
eigenvalue equation
\begin{equation}
H_{a}|\alpha\rangle=\lambda_{\alpha}|\alpha\rangle ,
\end{equation}
whose the eigenvalues and corresponding eigenstates, in the limit of a weak probe beam~$\Omega_{p}\ll \gamma_{1},\gamma_{2}$, are
\begin{eqnarray}
\lambda_{+} &=& +c^{2}\Omega_{R} ,\qquad \quad |+\rangle = s|0\rangle+c|2\rangle ,\nonumber\\
\lambda_{-} &=& -s^{2}\Omega_{R} ,\qquad \quad |-\rangle=s|2\rangle-c|0\rangle ,\nonumber\\
\lambda_{1} &=& -(\omega_{21}-\Delta),\quad |1\rangle,
\end{eqnarray}
where
\begin{eqnarray}
c^{2} =\frac{1}{2} +\frac{\Delta}{2\Omega_{R}} ,\quad
s^{2} =\frac{1}{2} -\frac{\Delta}{2\Omega_{R}} ,
\end{eqnarray}
and $\Omega_{R} = \sqrt{\Delta^{2}+4\Omega_{L}^{2}}$ is the detuned Rabi frequency of the driving field.

We now introduce the interaction between the dressed atom and the cavity field and work in the
bad cavity limit~\cite{gaox,Knight,Zhu3},  in which the cavity decay dominates over the coupling strengths $g_{1}, g_{2}$ and the atomic decay rates $\gamma_{1}$ and $\gamma_{2}$, i.e.
\begin{equation}
\kappa\gg{g_{1}},g_{2}\gg{\gamma_{1}},\gamma_{2} .
\end{equation}
Such a feature implies that the cavity mode response to the standard
vacuum reservoir is much faster than that produced by its
interaction with the atom. In other words, the cavity field forms a
finite bandwidth (Markovian) vacuum reservoir.

Working in the bad cavity limit, we can adiabatically eliminate the
cavity variables. This yields a master equation where the damping
terms have a structure dependent on the difference between the
cavity field and the dressed-atom transition frequencies. Details of
the adiabatic approximation have been presented in ref.~\cite{gaox}.
Here, we will apply such approach to study the linear and nonlinear
responses of the system to a weak probe field.

From the cavity-modified master equation, the equation of
motion for the atomic density matrix elements, written in the dressed state basis, are of the form
\begin{eqnarray}
\dot{\rho}_{--} &=& -R_{-+}\rho_{--}+R_{+-}\rho_{++}+R_{1-}\rho_{11}
+ s\left(x_{1}\tilde{\rho}_{-1}e^{i\Delta_{p}t}
+x_{1}^{\ast}\tilde{\rho}_{1-}e^{-i\Delta_{p}t}\right) \nonumber \\
&& +i\Omega_{p}c\left(\tilde{\rho}_{1-} - \tilde{\rho}_{-1}\right) , \label{e11}\\
\dot{\rho}_{11}&=& -(R_{1+}+R_{1-})\rho_{11}
- s\left(x_{2}\tilde{\rho}_{-1}e^{i\Delta_{p}t} + x_{2}^{\ast}\tilde{\rho}_{1-}e^{-i\Delta_{p}t}\right) \nonumber \\
&& +i\Omega_{p}\left[s\left(\tilde{\rho}_{1+} - \tilde{\rho}_{+1}\right) -c\left(\tilde{\rho}_{1-} - \tilde{\rho}_{-1}\right)\right] , \label{e12} \\
\dot{\tilde{\rho}}_{-1} &=& -\left[\Gamma_{-} +
i(\omega_{21}-\lambda_{-}-\Delta_{p})\right] \tilde{\rho}_{-1}
-s\left(x_{4}\rho_{11}+x_{2}^{\ast}\rho_{--}\right)e^{-i\Delta_{p}t} \nonumber \\
&& +i\Omega_{p}\left[s\rho_{-+} + c(\rho_{11}-\rho_{--})\right] , \label{e13} \\
\dot{\tilde{\rho}}_{1+}&=& -\left[\Gamma_{+}^{*} +
i(\omega_{21}-\lambda_{+}+\Delta_{p})\right]\tilde{\rho}_{1+}
-x_{2}e^{i\Delta_{p}t}\rho_{-+} \nonumber \\
&& +i\Omega_{p}\left[s\left(\rho_{11} - \rho_{++}\right) + c\rho_{-+}\right] , \label{e14}\\
\dot{\rho}_{-+} &=& -(\Gamma_{0}^{*}+i\Omega_{R}) \rho_{-+} -
x_{3}se^{-i\Delta_{p}t}\tilde{\rho}_{1+}
+i\Omega_{p}\left(s\tilde{\rho}_{-1} + c\tilde{\rho}_{1+}\right)
,\label{e15}
\end{eqnarray}
where
\begin{eqnarray}
x_{1} &=& (c^{2}-s^{2})\gamma_{12} +\frac{g_{1}g_{2}}{\kappa}\left(B_{0}-B_{3}^{\ast}\right) ,\nonumber\\
x_{2} &=& \gamma_{12} +\frac{g_{1}g_{2}}{\kappa}\left(B_{0}+B_{1}\right) ,\nonumber\\
x_{3} &=& (2cs+1)\gamma_{12}
+ \frac{g_{1}g_{2}}{\kappa}\left(B_{0}^{\ast}+2B_{4}+B_{3}\right) ,\nonumber\\
x_{4} &=& \gamma_{12} + \frac{g_{1}g_{2}}{\kappa}\left(B_{3}+B_{4}\right) ,
\end{eqnarray}
are the quantum interference terms,
\begin{eqnarray}
R_{+-} &=& 2c^{4}\left(\gamma_{2} +
\frac{g_{2}^{2}}{c^{4}\kappa}\left|B_{2}\right|^{2}\right) ,\quad
R_{-+} = 2s^{4}\left(\gamma_{2} + \frac{g_{2}^{2}}{s^{4}\kappa}\left|B_{1}\right|^{2}\right) ,\nonumber\\
R_{1-} &=& 2c^{2}\left(\gamma_{1} +
\frac{g_{1}^{2}}{c^{4}\kappa}\left|B_{4}\right|^{2}\right) ,\quad
R_{1+} = 2s^{2}\left(\gamma_{1} +
\frac{g_{1}^{2}}{s^{4}\kappa}\left|B_{3}\right|^{2}\right) ,
\end{eqnarray}
are the cavity modified damping rates between the dressed states,
\begin{eqnarray}
\Gamma_{0} &=& \gamma_{2}(1+2c^{2}s^{2})+ \frac{g_{2}^{2}}{\kappa}
\left[s^{2}\left(2B_{0}+2B_{0}^{\ast}+B_{1}\right)+c^{2}B_{2}\right] ,\nonumber\\
\Gamma_{-} &=& \gamma_{1}+\frac{g_{1}^{2}}{\kappa}\left(B_{3}^{\ast}
+B_{4}^{\ast}\right)
+s^{2}\left[\gamma_{2}+\frac{g_{2}^{2}}{\kappa}\left(B_{0}+B_{1}\right)\right] ,\nonumber\\
\Gamma_{+} &=& \gamma_{1}+\frac{g_{1}^{2}}{\kappa}\left(B_{3}^{\ast}
+B_{4}^{\ast}\right)
+c^{2}\left[\gamma_{2}+\frac{g_{2}^{2}}{\kappa}B_{2}\right]+s^{2}\frac{g_{2}^{2}}{\kappa}B_{0},
\end{eqnarray}
are the cavity modified damping rates of the coherence, with
\begin{eqnarray}
B_{0} &=& \frac{c^{2}\kappa}{\kappa+i\delta_{c}} ,\quad
B_{1} = \frac{s^{2}\kappa}{\kappa+i(\delta_{c}+\Omega_{R})} ,\quad
B_{2} = \frac{c^{2}\kappa}{\kappa+i(\delta_{c}-\Omega_{R})} ,\nonumber\\
B_{3} &=& \frac{s^{2}\kappa}{\kappa+i(\delta_{c}+\omega_{21}-\lambda_{-})} ,\quad
B_{4} = \frac{c^{2}\kappa}{\kappa+i(\delta_{c}+\omega_{21}-\lambda_{+})} ,\label{e18}
\end{eqnarray}
and
\begin{eqnarray}
\tilde{\rho}_{1-} &=& \rho_{1-}e^{i\Delta_{p}t} ,\quad
\tilde{\rho}_{-1} = \rho_{-1}e^{-i\Delta_{p}t} ,\nonumber \\
\tilde{\rho}_{1+} &=& \rho_{1+}e^{i\Delta_{p}t} ,\quad
\tilde{\rho}_{+1} = \rho_{+1}e^{-i\Delta_{p}t} ,
\end{eqnarray}
are the dressed atom coherence in a rotating frame oscillating with
frequency $\Delta_{p}$. Equations (\ref{e11})$-$\ref{e15}) are valid
for any value of the cavity detuning $\delta_{c}$, and the upper
levels splitting comparable to half of the Rabi frequency, i.e. for
$\omega_{12}-\lambda_{+}\sim \gamma_{i}$. Physically, the
approximation of $\omega_{12}-\lambda_{+}\sim \gamma_{i}$
corresponds to the case when the probe level $\ket 1$ is degenerate
or nearly degenerate with respect to the dressed state $\ket -$. In
this case, the resultant degeneracy gives rise to maximal quantum
interference effects. Moreover, under this approximation,
$B_{4}\approx B_{0}$ and $B_{3}\approx B_{1}$.

The parameters appearing in the equations of motion have simple
physical interpretations. The parameters $x_{i}$  are quantum
interference terms. They contain contributions of both,
spontaneously generated and cavity effects, which clearly illustrate
an analogy between the cavity engineered and the spontaneously
induced coherence~\cite{Zhu,gaox,fs04}.  Thus, the cavity with large
decay rate strongly  enhances quantum interference effects.

The parameters $R_{ij}$ represent the transition rates between the
dressed states of the system and~$\Gamma_{i}$ are the damping rates
of the coherence. Note that the parameters are dependent on the Rabi
frequency of the driving field and are resonant when the cavity
frequency is tuned to $\delta_{c} =0, \pm \Omega_{R},
\lambda_{\pm}-\omega_{21}$. It means that spontaneous emission and
quantum interference dominate at five frequencies. The sensitivity
of the coefficients on~$\delta_{c}$ is known in the literature as
the Purcell effect. Thus, in the system considered here, both the
Purcell and the cavity-induced quantum interference effects play an
important role in the dynamics and properties of the system.

One can notice from (\ref{e11})$-$(\ref{e15}) that the coefficients in the differential equations are dependent on time. In fact, there is no reference frame in which the coefficients would be time independent. It is clear that the time dependence of the coefficients is brought here by the interference terms. As the result of the time dependence, special mathematical techniques must be employed to solve the set of the equations of motion. In the next section, we will solve the set of equations for the steady state density matrix elements using the Floquet technique.

\section{Linear and nonlinear (Kerr) susceptibilities}\label{ch3}

Our purpose of this paper is to demonstrate that the combined effect
of the Purcell and the cavity-induced quantum interference phenomena
can create giant linear and nonlinear susceptibilities in the
three-level system. Note that the cavity-induced quantum
interference effects are more flexible to the parameters than those
induced by the spontaneously generated coherence. The latter depend
solely on the angle between the dipole moments of the two atomic
transitions. The former depend on the Rabi frequency of the driving
field, damping rates of the atomic transitions, and the detunings of
the fields from their resonances. This makes the cavity system more
practical for creation of  quantum interference effects than that
induced by spontaneously created coherence.

It is well known that the response of the atomic medium to the probe
field is governed by its polarization $P$, which can be expressed in
terms of the complex susceptibility $\chi$ or related to the
elements of the density matrix of the system as
\begin{equation}
P = \varepsilon_{0}(E_{p}\chi+E_{p}^{*}\chi^{*}) = 2N_{a}(\mu_{01}\rho_{10}+\mu_{10}\rho_{01}) ,
\end{equation}
where $N_{a}$ is the number density of the atoms, and
$\rho_{10}=s\rho_{1+}-c\rho_{1-}$ is the atomic coherence on the
probed transition. The task then is to determine the atomic
coherence $\rho_{10}$, or equivalently $\rho_{1+}$ and $\rho_{1-}$,
which could be found by solving the set of
equations~(\ref{e11})$-$(\ref{e15}). In the stationary limit
$t\rightarrow \infty$, we may set all of the derivatives to zero and
obtain a set of algebraic equations for the density matrix elements.
However, the set of equations retains the time dependence through
the factors $\exp{(\pm i\Delta_{p}t)}$. Therefore, to solve the
system of equations (\ref{e11})$-$(\ref{e15}), we employ the Floquet
method by expressing the density matrix elements as Fourier series
in terms of amplitudes that oscillate at the probe detuning and its
harmonics. As we are interested in the response of the system to a
weak probe field, we also make an expansion of the density matrix
elements in terms of the powers of the probe field. These two
decompositions combined together are given by the
relation~\cite{ever}
\begin{eqnarray}
\rho_{jk}=\sum_{m=0}^{+\infty}\sum_{n=-\infty}^{+\infty}
\lambda^{m}(\rho_{jk})_{m}^{n}\ {\rm e}^{in\Delta_{p}t} ,
\label{e21}
\end{eqnarray}
where the expansion in the powers of $\Omega_{p}$ is given in terms
of a dimensionless parameter~$\lambda$ that can take on values
ranging continuously from zero (no perturbation) to one (the full
perturbation). 

Since the atomic coherence on the probe transition oscillates as $\exp(i\Delta_{p}t)$, the stationary properties of the first and third-order susceptibilities are determined by the harmonics
$(\rho_{10})^{-1}_{1}$ and $(\rho_{10})^{-1}_{3}$, respectively.
Therefore according to Eq.~(10) the susceptibilities $\chi^{(1)}$ and~$\chi^{(3)}$ can be expressed in terms of the first and third order coherence of the probe transition as
\begin{eqnarray}
\chi^{(1)} &=&
\frac{-2N_{a}|\vec{\mu}_{13}|^{2}}{\hbar\varepsilon_{0}\Omega_{p} }
\left[s(\rho_{1+})_{1}^{-1}-c(\rho_{1-})_{1}^{-1}\right] , \label{e22}\\
\chi^{(3)} &=&
\frac{-2N_{a}|\vec{\mu}_{13}|^{4}}{3\hbar^{3}\varepsilon_{0}\Omega_{p}^3} 
\left[s(\rho_{1+})_{3}^{-1}-c(\rho_{1-})_{3}^{-1}\right] .\label{e23}
\end{eqnarray}
The linear and nonlinear susceptibilities can be conveniently expressed in the form
\begin{eqnarray}
\chi^{(k)} = -\frac{2N_{a}|\vec{\mu}_{13}|^{k+1}}{\left(\sqrt{3}\right)^{k-1}\hbar^{k}\varepsilon_{0}} 
\left[{\rm Re}\chi^{(k)} +i{\rm Im}\chi^{(k)}\right] ,\quad k=1,3 ,\label{e24}
\end{eqnarray}
where we have introduced the normalized real and imaginary parts of
$\chi^{(k)}$ that determine the index of refraction and the absorption coefficient, respectively. Evidently,
the normalized parts of $\chi^{(k)}$ are independent of the probe field strength~$\Omega_p$. This allows the susceptibilities to be arbitrary large since the only approximation made here is an assumption of weak probe beam strengths. 
Note that ${\rm Im}\chi^{(k)}=0$ implies lossless propagation of the probe field,
${\rm Re}\chi^{(1)}\neq 0$ implies linear refraction of the probe
beam, and ${\rm Re}\chi^{(3)}\neq 0$ implies nonlinear intensity
dependent (Kerr) refraction.

Upon substitution of (\ref{e21}) into (\ref{e11})$-$(\ref{e15}) and
after comparing terms of the same powers in $n\Delta_{p}$, we obtain
an infinite set of equations for the Fourier harmonics with time
independent coefficients. Despite of the complexity, the system of
the coupled equations is easily solved for the steady state by an
iteration in terms of the powers of the probe field amplitude. The
analytical iterative solution for the~$(n,m)$ order harmonics of the
coherence appearing in (\ref{e22}) and (\ref{e23}) are of the form
\begin{eqnarray}
(\rho_{1+})_{m}^{n} &=& -i\Omega_{p}\frac{(\Gamma_{2}^{*}+in\Delta_{p})[s(\rho_{-1})_{m-1}^{n+1}+ c(\rho_{1+})_{m-1}^{n-1}]}{(\Gamma_{1}^{*}+in\Delta_{p})(\Gamma_{2}^{*}+in\Delta_{p})-x_{2}x_{3}^{*}} \nonumber \\
&&+
i\Omega_{p}\frac{x_{2}\left\{s[(\rho_{11})_{m-1}^{n+1}-(\rho_{++})_{m-1}^{n+1}]
+c(\rho_{-+})_{m-1}^{n+1}\right\}
}{(\Gamma_{1}^{*}+in\Delta_{p})(\Gamma_{2}^{*}+in\Delta_{p})
-x_{2}x_{3}^{*}} ,\label{e25} \\
(\rho_{1-})_{m}^{n} &=&
-i\Omega_{p}\frac{c[(\rho_{11})_{m-1}^{n+1}-(\rho_{--})_{m-1}^{n+1}]
+s(\rho_{+-})_{m-1}^{n+1}}{\Gamma_{3}^{*}-in\Delta_{p}} \nonumber \\
&&-\frac{s\left[x_{4}^{*}(\rho_{11})_{m}^{n}
+x_{2}(\rho_{--})_{m}^{n}\right]}{\Gamma_{3}^{*}-in\Delta_{p}} ,\label{e26}
\end{eqnarray}
where the analytical solutions for the auxiliary harmonics are quite lengthy and are listed in the Appendix. It follows from the explicit solutions (\ref{e25}) and (\ref{e26}) that the magnitudes of the 
harmonics are proportional to $\Omega_{p}^{m}$, which ensures that their magnitudes are small even if the normalized susceptibilities ${\rm Re}\chi^{(k)}$ and ${\rm Im}\chi^{(k)}$ are large, since the probe beam strength is considered here to be weak, $\Omega_{p}\ll \gamma_{1},\gamma_{2}$. This justifies the power expansion~(\ref{e21}).

While (\ref{e25}) and (\ref{e26}) constitute an analytical solution to the susceptibility of the atomic medium, their form is algebraically complicated and there is little to be gained from a detailed dissection of these results. Therefore, we will perform numerical analysis.

\section{Discussion of the results}\label{ch4}

We now proceed to perform detailed analysis of the the linear and nonlinear susceptibilities by
graphically displaying the real and imaginary parts of $\chi^{(1)}$ and $\chi^{(3)}$ for a wide range of the important parameters.  We are particularly interested in the possibility of creation of giant Kerr nonlinearities accompanied by zero linear and nonlinear absorptions. In what follows, we assume for simplicity that the driving laser field is on resonance with the atomic transition $|2\rangle\leftrightarrow|0\rangle$, i.e.,~$\Delta=0$.
\begin{figure}[h]
\begin{center}
\includegraphics[width=10cm]{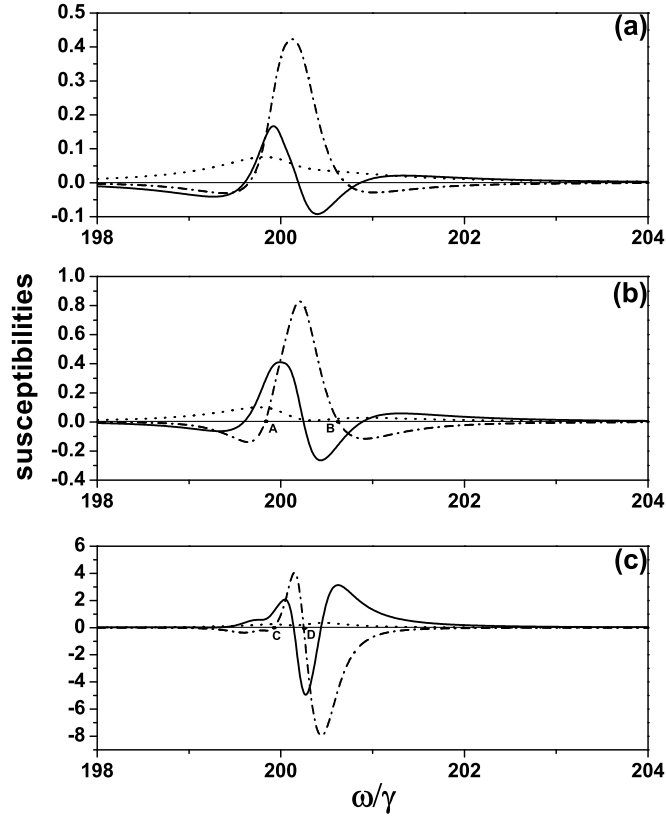}
\end{center}
\caption{The Kerr nonlinearity ${\rm Re}\chi^{(3)}$ (solid line), the linear ${\rm Im}\chi^{(1)}$ (dotted line) and nonlinear ${\rm Im}\chi^{(3)}$ (dashed-dotted line) absorption coefficients plotted as a function of the probe detuning $\omega/\gamma = (\omega_{p}-\omega_{1})/\gamma $ for  $\kappa=100\gamma$,
$g_{2}=15\gamma$, $g_{1}=5\gamma$, $\gamma_{1}=\gamma_{2}=0.1\gamma$,
$\omega_{21}=\Omega_{L}=200\gamma$, and (a) $\delta_{c} =0$, (b) $\delta_{c}=50\gamma$, (c) $\delta_{c}=200\gamma$.}
\label{fig2}
\end{figure}

In Fig.~\ref{fig2}, we illustrate the variation of the real and
imaginary parts of $\chi^{(3)}$ and the imaginary part of
$\chi^{(1)}$ with the probe field detuning
$\omega=\omega_{p}-\omega_{1}$ from resonance with the transition
$\ket 0\leftrightarrow \ket 1$. We choose the Rabi frequency of the
driving field such that $\Omega_{L}=\omega_{21}$. In this particular
case, the dressed state $\ket -$ and the probe state $\ket 1$ are
degenerated in the energies,  which is the maximal quantum
interference configuration. All of the parameters are measured in
units of the damping rate $\gamma$ through out these figures. Part
(a) of the figure shows the susceptibilities for $\delta_{c} =0$.
This corresponds to the cavity field tuned to the central component
of the dressed transitions. We see that the susceptibilities exhibit
resonance structures in the vicinity of the frequency
$\omega=200\gamma$. As we have already mentioned, at this frequency
the quantum interference is maximal. The nonlinear (Kerr)
susceptibility is enhanced, but at the same time the linear and
nonlinear absorptions are large. Even at the frequency where the
nonlinear absorption, Im$\chi^{(3)}$, vanishes, the linear
absorption Im$\chi^{(1)}$ is large with the magnitude comparable to
the magnitude of the Kerr nonlinearity. This is not desirable for a
practical application since the probe beam could be completely
absorbed over a short distant of propagation inside the atomic
medium. Therefore, we now proceed to check if one could achieve a
large Kerr nonlinearity accompanied by vanishing linear and
nonlinear absorption by varying parameters of the system. A close
inspection of the analytical expressions (\ref{e25}) and (\ref{e26})
shows that the absorption rate of the probe beam depends on the
difference $\rho_{11}-\rho_{\pm,\pm}$ between the populations of the
lower and upper levels of the probe transition which, on the other
hand, depends on the detuning $\delta_{c}$. Thus, we expect that the
transparency of the propagation of the probe beam could be improved
by applying the Purcell effect, i.e. by varying the detuning
$\delta_{c}$ to match the cavity frequency with the frequency of one
of the Rabi sidebands of the driven transition.

Parts (b) and (c) of the figure show how the susceptibilities are modified when the cavity detuning
$\delta_{c}$ is varied. There a few significant changes observed in the behavior of the susceptibility. Firstly, the Kerr nonlinearity becomes enhanced by few orders in magnitude when the detuning
$\delta_{c}$  approaches the value $\delta_{c}=200\gamma$, corresponding to the tuning of the cavity field to the Rabi sideband of the driven transition. Secondly, the Kerr nonlinearity varies rapidly with the probe frequency. However, the most important change in the behavior of the susceptibility is that at the frequency, indicated by a dot~D, where the Kerr nonlinearity is maximal, the nonlinear absorption vanishes completely and the linear absorption is negligibly small. In other words, the system is transparent for the probe beam at the frequency where the Kerr nonlinearity is maximal. We may conclude that by tuning the cavity field to one of the Rabi sidebands, one can achieve a giant Kerr nonlinearity accompanied by vanishing absorption.

\begin{figure}[h]
\begin{center}
\includegraphics[width=10cm]{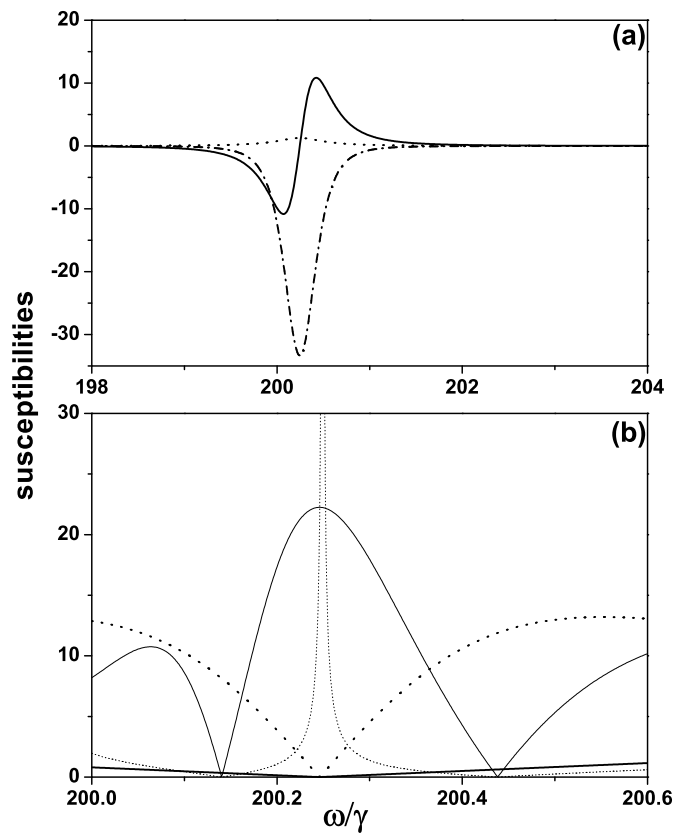}
\end{center}
\caption{(a) The Kerr nonlinearity ${\rm Re}\chi^{(3)}$ (solid line), the linear ${\rm Im}\chi^{(1)}$ (dotted line) and nonlinear ${\rm Im}\chi^{(3)}$ (dashed-dotted line) absorption coefficients plotted as a function of the probe detuning $\omega/\gamma = (\omega_{p}-\omega_{1})/\gamma $ for the same parameters as in Fig.~\ref{fig2}(c) but $\omega_{21}=250\gamma$. The bottom part  (b) presents the ratios ${\rm Re}\chi^{(3)}/{\rm Im}\chi^{(1)}$ (solid lines) and ${\rm Re}\chi^{(3)}/{\rm Im}\chi^{(3)}$ (dotted  lines) plotted as a function of  $\omega/\gamma = (\omega_{p}-\omega_{1})/\gamma $ for the same parameters as in Fig.~\ref{fig2}(c) but two different values of $\omega_{12}$: $\omega_{21}=200\gamma$ (thick solid and dotted lines) and $\omega_{21}=250\gamma$ (thin solid and dotted lines).}
\label{fig3}
\end{figure}

We have seen that a combination of the maximal quantum interference
and the Purcell effect is crucial for creation of a giant Kerr
nonlinearity accompanied by vanishing absorption. To illustrate the
importance of maintaining the maximal quantum interference, we now
slightly detune the dressed state $\ket -$ from the probed atomic
state $\ket 1$, so that the states become non-degenerate. It is well
known, that quantum interference effects degrade when interfering
energy states are non-degenerate. Let us see how this can affect the
Kerr nonlinearity and the transparency of the atomic medium. In
Fig.~\ref{fig3}, we plot the imaginary parts of $\chi^{(1)}$ and
$\chi^{(3)}$ and the real part of $\chi^{(3)}$ for the same
parameters as in Fig.~\ref{fig2}(c), but with
$\omega_{21}=250\gamma$. In this case, the state $\ket -$ is detuned
from the state $\ket 1$ by $50\gamma$, that is the cavity field is
detuned from the dressed atom frequencies. It is easy to see from
(\ref{e18}) that the effect of detuning the cavity field from the
dressed atom frequencies is to reduce the magnitude of quantum
interference terms. Part (a) of the figure shows that the linear
absorption is small at all frequencies, but within the region when
the Kerr nonlinearity is enhanced, the nonlinear absorption is very
large. Thus, the atomic medium becomes highly absorbing for the
probe beam when the quantum interference effects are reduced.

To illustrate further the effectiveness of the enhancement of the
Kerr nonlinearity by quantum interference, we plot in part (b) of
the figure the ratios ${\rm Re}\chi^{(3)}/{\rm Im}\chi^{(3)}$ and
${\rm Re}\chi^{(3)}/{\rm Im}\chi^{(1)}$ for the presence and the
absence of quantum interference. We see that at the frequency
$\omega =200.25\gamma$ the ratios are maximal in the presence of
quantum interference and vanish completely in the absence of quantum
interference. Note that the maxima of the ratios occur at
frequencies slightly shifted from the resonance $\omega=200\gamma$.
This is because after adiabatically eliminating the cavity field
operators in the bad cavity limit, the remaining cavity effects are
not only to affect the atomic damping rates but also to induce a
small energy shifts for the levels $|\pm\rangle$ and $|1\rangle$.
We may conclude that the enhanced Kerr nonlinearity with relatively
vanishing linear and nonlinear absorptions is a signature of the
cavity-induced quantum interference effects.
\begin{figure}[h]
\begin{center}
\includegraphics[width=10cm]{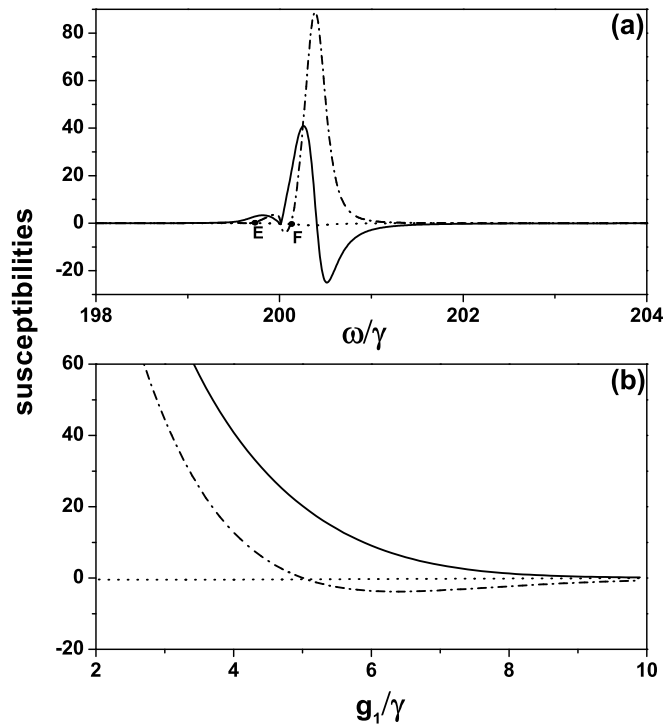}
\end{center}
\caption{(a) The Kerr nonlinearity ${\rm Re}\chi^{(3)}$ (solid line), the linear ${\rm Im}\chi^{(1)}$ (dotted line) and nonlinear ${\rm Im}\chi^{(3)}$ (dashed-dotted line) absorption coefficients plotted as a function of the probe detuning $\omega/\gamma = (\omega_{p}-\omega_{1})/\gamma $ for the same parameters as in Fig.~\ref{fig2}(c) but a very small damping rate on the probe transition $\gamma_{1}=0.001\gamma$. The bottom part (b) presents the Kerr nonlinearity ${\rm Re}\chi^{(3)}$ (solid line), the linear ${\rm Im}\chi^{(1)}$ (dotted line) and nonlinear ${\rm Im}\chi^{(3)}$ (dashed-dotted line) absorption coefficients plotted as a function of the coupling constant $g_{1}/\gamma$  for the same parameters as in Fig.~\ref{fig2}(c) but $\gamma_{1}=0.001\gamma$ and $\omega=200.122\gamma$.}
\label{fig4}
\end{figure}

We now proceed to check the importance of other parameters of the system such as the atomic
decay rate $\gamma_{1}$ and the cavity field-atom coupling constant $g_{1}$.
Figure~\ref{fig4}(a) illustrates the susceptibility for the same parameters as in Fig.~\ref{fig2}(c), but
$\gamma_{1}=0.001\gamma$. It is evident that at this small damping rate, the linear absorption is zero at all frequencies, while the nonlinear absorption vanishes at two frequencies, indicated by dots E and F.
At these frequencies the Kerr nonlinearity is large. In particular, at the point F, the Kerr nonlinearity is about one order higher in magnitude than that observed in Fig.~\ref{fig2}(c) for
$g_{1}=5\gamma$. Evidently, the Kerr nonlinearity can be enhanced and the linear and nonlinear absorptions kept zero by a proper choosing of the atomic decay rate on the probe transition.

The dependence of the Kerr nonlinearity and the absorption
coefficients on the coupling constant $g_{1}$ is illustrated in
Fig.~\ref{fig4}(b). We show the imaginary parts of the
susceptibilities $\chi^{(1)}$ and $\chi^{(3)}$ and the real part of
$\chi^{(3)}$ as a function of $g_{1}$ for the probe detuning $\omega
= 200.122\gamma$ corresponding to the position of the maximum of
${\rm Re}\chi^{(3)}$. It is interesting to note that the linear
absorption rate is zero independent of $g_{1}$, while the nonlinear
absorption varies from positive to negative and vanishes at
$g_{1}=5.0\gamma$. This shows that one can varies the magnitudes of
the Kerr nonlinearity and the absorption coefficients by a proper
setting of the coupling constant.

We close this section by a brief analysis of the adiabatic approximation and the range of the parameters used in our analytical treatment of the nonlinear dynamics of the system. One could object that the values of the parameters selected for plotting the figures are not in the range to fulfill the bad cavity limit of $\kappa \gg g_{1},g_{2}$. In the first instance, we solve the master equation (\ref{e1}) numerically, using the quantum optics toolbox for Matlab~\cite{she}, for the steady-state values of the zeroth harmonics of the populations and coherence of the dressed states that determine the susceptibility of the system. We use the same values for the parameters as in Fig.~\ref{fig2}(c), and the analytical and numerical results are listed in the table. It is evident that the discrepancies between the values of the density matrix elements obtained by the approximate solutions and corresponding exact numerical results are negligibly small.\\

\begin{tabular}[h]{|l|l|l|}
\hline
$\rho_{ij}$ & Analytical solution           & Exact numerical solution\\ \hline
 $\rho_{11}$ &0.2072        & 0.2082 \\
 $\rho_{++}$ &0.2409        & 0.2375\\
 $\rho_{--}$ &0.5520             & 0.5543\\
$\rho_{-1}$ &-0.0086 -0.1749$i$ & -0.0093-0.1755$i$\\ \hline
\end{tabular}\\

\begin{figure}[h]
\begin{center}
\includegraphics[width=10cm]{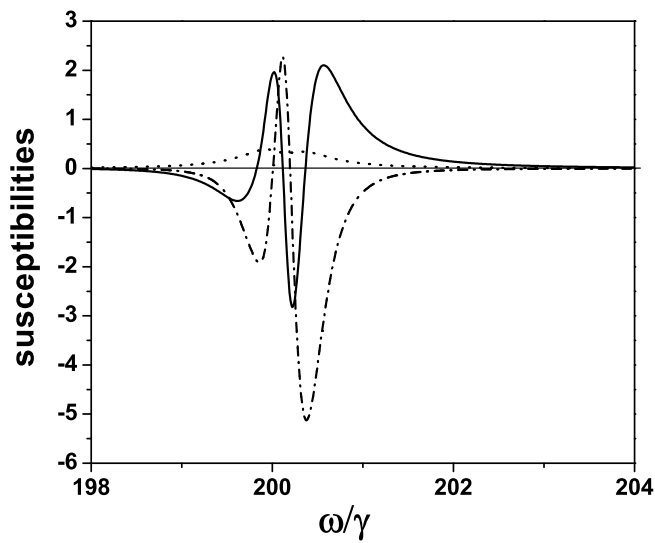}
\end{center}
\caption{(a) The Kerr nonlinearity ${\rm Re}\chi^{(3)}$ (solid line), the linear ${\rm Im}\chi^{(1)}$ (dotted line) and nonlinear ${\rm Im}\chi^{(3)}$ (dashed-dotted line) absorption coefficients plotted as a function of the probe detuning $\omega/\gamma = (\omega_{p}-\omega_{1})/\gamma $ for the same parameters as in Fig.~\ref{fig2}(c) but $\kappa =200\gamma$.}
\label{fig5}
\end{figure}

In the second, we plot in Fig.~\ref{fig5} the Kerr nonlinearity
together with the linear and nonlinear absorption coefficients for
the same parameters as in Fig.~\ref{fig2}(c) but a significantly
larger cavity damping rate, $\kappa =200\gamma$. We observe that the
effects are qualitatively the same as those predicted for
$\kappa=100\gamma$. The Kerr nonlinearity attains maximal value at
frequencies where the linear and nonlinear absorption are
negligible. The only difference is in the numerical values of the
magnitudes of the real and imaginary parts of the susceptibility.

\section{Summary}\label{ch5}

We have studied the linear and nonlinear responses to a weak probe
beam of a three-level atom coupled to a single-mode cavity and
driven by a strong laser field. Working in the bad cavity limit, we
derived analytical expressions for the linear and nonlinear (Kerr)
susceptibilities. We have found that the joint effect of quantum
interference and the Purcell effect can lead to a giant Kerr
nonlinearity of the atomic medium accompanied by vanishing
absorption. We have shown that the presence of maximal quantum
interference is crucial for creation of the complete transparency of
the atomic medium. The role of the significant parameters of the
system has been discussed in details. We have shown that the
creation of a giant Kerr nonlinearity accompanied by vanishing
absorption can be easily accomplished by a proper setting of the
atomic decay rates or by a proper adjusting of the cavity field-atom
coupling constants.

\section{Acknowledgments}

The authors acknowledge financial support from  the National Natural
Science Foundation of China (under Grants Nos. 10674052 and
60878004), the Ministry of Education under project NCET (under grant
no. NCET-06-0671) and SRFDP(under grant no. 200805110002) , and the
National Fundamental Research Program of China (under Grant No.
2005CB724508).


\section*{Appendix}

In the appendix we present the analytical iterative solutions for the steady-state values of the Fourier harmonics of the density matrix elements involved in the calculation of the linear and nonlinear susceptibilities. The $(n,m)$ harmonics are of the form
\begin{eqnarray}
(\rho_{-1})_{m}^{n} &=&
i\Omega_{p}\frac{c[(\rho_{11})_{m-1}^{n-1}-(\rho_{--})_{m-1}^{n-1}]
+
s(\rho_{-+})_{m-1}^{n-1}}{(\Gamma_{3}+in\Delta_{p})} \nonumber\\
&& +s\frac{x_{4}(\rho_{11})_{m}^{n}
+x_{2}^{\ast}(\rho_{--})_{m}^{n}}{(\Gamma_{3}+in\Delta_{p})} ,\nonumber \\
(\rho_{-+})_{m}^{n}&=& i\Omega_{p}\frac{x_{3}^{\ast}[s(\rho_{-1})_{m-1}^{n+1}+ c(\rho_{1+})_{m-1}^{n-1}] }{(\Gamma_{1}^{*}+in\Delta_{p})(\Gamma_{2}^{*}+in\Delta_{p})-x_{2}x_{3}^{*}} \nonumber\\
&&+i\Omega_{p}\frac{(\Gamma_{1}^{*}+in\Delta_{p})
\left\{s[(\rho_{++})_{m-1}^{n+1}-(\rho_{11})_{m-1}^{n+1}]
-c(\rho_{-+})_{m-1}^{n+1}\right\}}{(\Gamma_{1}^{*}+in\Delta_{p})(\Gamma_{2}^{*}+in\Delta_{p})-x_{2}x_{3}^{*}} ,\nonumber\\
(\rho_{+1})_{m}^{n} &=& i\Omega_{p}\frac{(\Gamma_{2}+in\Delta_{p})\left[s(\rho_{1-})_{m-1}^{n-1} +c(\rho_{+1})_{m-1}^{n+1}\right]}{(\Gamma_{1}+in\Delta_{p})(\Gamma_{2}+in\Delta_{p})-x_{2}^{*}x_{3}} \nonumber\\
&&
-i\Omega_{p}\frac{x_{2}^{*}\left\{s[(\rho_{++})_{m-1}^{n-1}-(\rho_{11})_{m-1}^{n-1}]
+c(\rho_{+-})_{m-1}^{n-1}\right\}}{(\Gamma_{1}+in\Delta_{p})(\Gamma_{2}+in\Delta_{p})-x_{2}^{*}x_{3}} ,\nonumber\\
(\rho_{+-})_{m}^{n} &=& -i\Omega_{p}\frac{x_{3}\left[s(\rho_{1-})_{m-1}^{n-1} +c(\rho_{+1})_{m-1}^{n+1}\right]   }{(\Gamma_{1}+in\Delta_{p})(\Gamma_{2}+in\Delta_{p}) - x_{2}^{*}x_{3}} \nonumber\\
&&+ i\Omega_{p}\frac{(\Gamma_{1}+in\Delta_{p})\left\{s[(\rho_{++})_{m-1}^{n-1}-(\rho_{11})_{m-1}^{n-1}]+c(\rho_{+-})_{m-1}^{n-1}\right\} }{(\Gamma_{1}+in\Delta_{p})(\Gamma_{2}+in\Delta_{p}) - x_{2}^{*}x_{3}} , \nonumber \\
(\rho_{11})_{m}^{n} &=& \frac{-H_{4}U_{m-1}^{n\pm 1} -H_{1}W_{m-1}^{n\pm 1}}{H_{1}H_{3}+H_{2}H_{4}}, \nonumber \\
(\rho_{--})_{m}^{n} &=& \frac{H_{2}U_{m-1}^{n\pm 1}
-H_{3}W_{m-1}^{n\pm 1}}{H_{1}H_{3}+H_{2}H_{4}},
\end{eqnarray}
where
\begin{eqnarray}
\Gamma_{1} &=& \Gamma_{0} + i\Omega_{R} ,\quad 
\Gamma_{2} = \Gamma_{+} - i(\lambda_{+}-\omega_{21}) ,\quad 
\Gamma_{3} = \Gamma_{-} - i(\lambda_{+}-\omega_{21}) ,\nonumber\\
H_{1} &=& (R_{-+}+R_{+-}+in\Delta_{p})
+\frac{sx_{1}x_{2}^{\ast}}{\Gamma_{3}+in\Delta_{p}}
+\frac{sx_{1}^{*}x_{2}}{\Gamma_{3}^{*}+in\Delta_{p}} , \nonumber\\
H_{2} &=& (R_{+-}-R_{1-})+\frac{sx_{1}x_{4}}{\Gamma_{3}+in\Delta_{p}} + \frac{sx_{1}^{*}x_{4}^{\ast}}{\Gamma_{3}^{*}+in\Delta_{p}} ,\nonumber\\
H_{3}&=& (R_{1+}+R_{1-}+in\Delta_{p})
-\frac{sx_{2}x_{4}}{\Gamma_{3}+in\Delta_{p}}
-\frac{sx_{2}^{*}x_{4}^{\ast}}{\Gamma_{3}^{*}+in\Delta_{p}} ,\nonumber\\
H_{4}&=&\frac{s|x_{2}|^{2}}{\Gamma_{3}+in\Delta_{p}}+\frac{s|x_{2}|^{2}}{\Gamma_{3}^{*}+in\Delta_{p}},
\end{eqnarray}
and
\begin{eqnarray}
U_{m-1}^{n\pm 1} &=& -i\Omega_{p}\left[c(\rho_{1-})_{m-1}^{n-1} - c(\rho_{-1})_{m-1}^{n+1}\right]\nonumber \\
&&-i\Omega_{p}\frac{x_{1}\left\{s(\rho_{-+})_{m-1}^{n-1}
-c\left[(\rho_{--})_{m-1}^{n-1}-(\rho_{11})_{m-1}^{n-1})\right]\right\}}{(\Gamma_{3}+in\Delta_{p})} \nonumber\\
&&+i\Omega_{p}\frac{x_{1}^{*}\left\{s(\rho_{+-})_{m-1}^{n+1}
-c\left[(\rho_{--})_{m-1}^{n+1}-(\rho_{11})_{m-1}^{n+1})\right]\right\}}{(\Gamma_{3}^{*}+in\Delta_{p})}
,
\nonumber \\
W_{m-1}^{n\pm 1} &=&
-i\Omega_{p}\left[s(\rho_{1+})_{m-1}^{n-1}-c(\rho_{1-})_{m-1}^{n-1}
-s(\rho_{+1})_{m-1}^{n+1} +c(\rho_{-1})_{m-1}^{n+1}\right] \nonumber\\
&&+i\Omega_{p}\frac{x_{2}\left\{s(\rho_{-+})_{m-1}^{n-1}-c\left[(\rho_{--})_{m-1}^{n-1}
-(\rho_{11})_{m-1}^{n-1})\right]\right\}}{(\Gamma_{3}+in\Delta_{p})}\nonumber\\
&&-i\Omega_{p}\frac{x_{2}^{*}\left\{s(\rho_{+-})_{m-1}^{n+1}-c\left[(\rho_{--})_{m-1}^{n+1}
-(\rho_{11})_{m-1}^{n+1})\right]\right\}}{(\Gamma_{3}^{*}+in\Delta_{p})}.
\end{eqnarray}

\end{document}